\documentclass[a4paper,fleqn]{cas-dc}

\usepackage[numbers]{natbib}

\def\tsc#1{\csdef{#1}{\textsc{\lowercase{#1}}\xspace}}
\tsc{WGM}
\tsc{QE}
\tsc{EP}
\tsc{PMS}
\tsc{BEC}
\tsc{DE}

\begin{document}
\let\WriteBookmarks\relax
\def\floatpagepagefraction{1}
\def\textpagefraction{.001}
\shorttitle{Graphical Granulometry}
\shortauthors{I Rehberg et~al.}

\title [mode = title]{Graphical Magnetogranulometry of EMG909}                      
\author{Ingo Rehberg}[orcid=0000-0001-7511-2910]
\author{Reinhard Richter}[orcid=0000-0002-6380-5477]
\author{Stefan Hartung}
\address{Experimentalphysik V, Universit\"at Bayreuth, 95440 Bayreuth, Germany}

\begin{abstract}
The magnetization curve of the commercially available ferrofluid EMG909 is measured. It can adequately be described by a superposition of four Langevin terms. The effective dipole strength of the magnetic particles in this fluid is subsequently obtained by a graphical rectification of the magnetization curve based on the inverse Langevin function. The method yields the arithmetic and the harmonic mean of the magnetic moment distribution function, and a guess for the geometric mean and the relative standard deviation. It has the advantage that it does not require a prejudiced guess of the distribution function of the poly-disperse suspension of magnetic particles.\end{abstract}
\maketitle
\section{Introduction}
"Die krumme Linie kennt kein gr\"o{\ss}eres Wunder, als die gerade. Aber nicht umgekehrt." (The bent line does not know a greater marvel than the straight line. But not the other way round.) This statement from Friedrich Hebbel might be easier to justify from an aesthetic than from a mathematical point of view, but can be considered as our guideline in the pursuit of straightening the typical S-shape of magnetization curves to bring out their individual and specific characteristics more clearly. For that purpose, taking the inverse Langevin function of the magnetization seems to be the natural approach \cite{Rehberg2019}. While that rectification effort is expected to work exactly for monodisperse ferrofluids, the outcome for polydisperse fluids is slightly more complicated than just a straight line. However, it turns out to be very useful: It serves to provide the arithmetic and the harmonic mean of the particle distribution function. 

In this contribution, we apply that method of "graphical magnetogranulometry" \cite{Rehberg2019} to the magnetization curve of the ferrofluid EMG909, commercially available from Ferro\-tec Co. This is of special interest within these proceedings, because this fluid had been chosen for the investigation of the magnetically stabilized Kelvin-Helmholtz instability, which was presented at ICMF 2019 \cite{ICMF2019}, and is described in Ref.\,\cite{Richter2020}. 
\section{Magnetization of EMG909}
\begin{figure}
\centering
\includegraphics[width=0.98\linewidth]{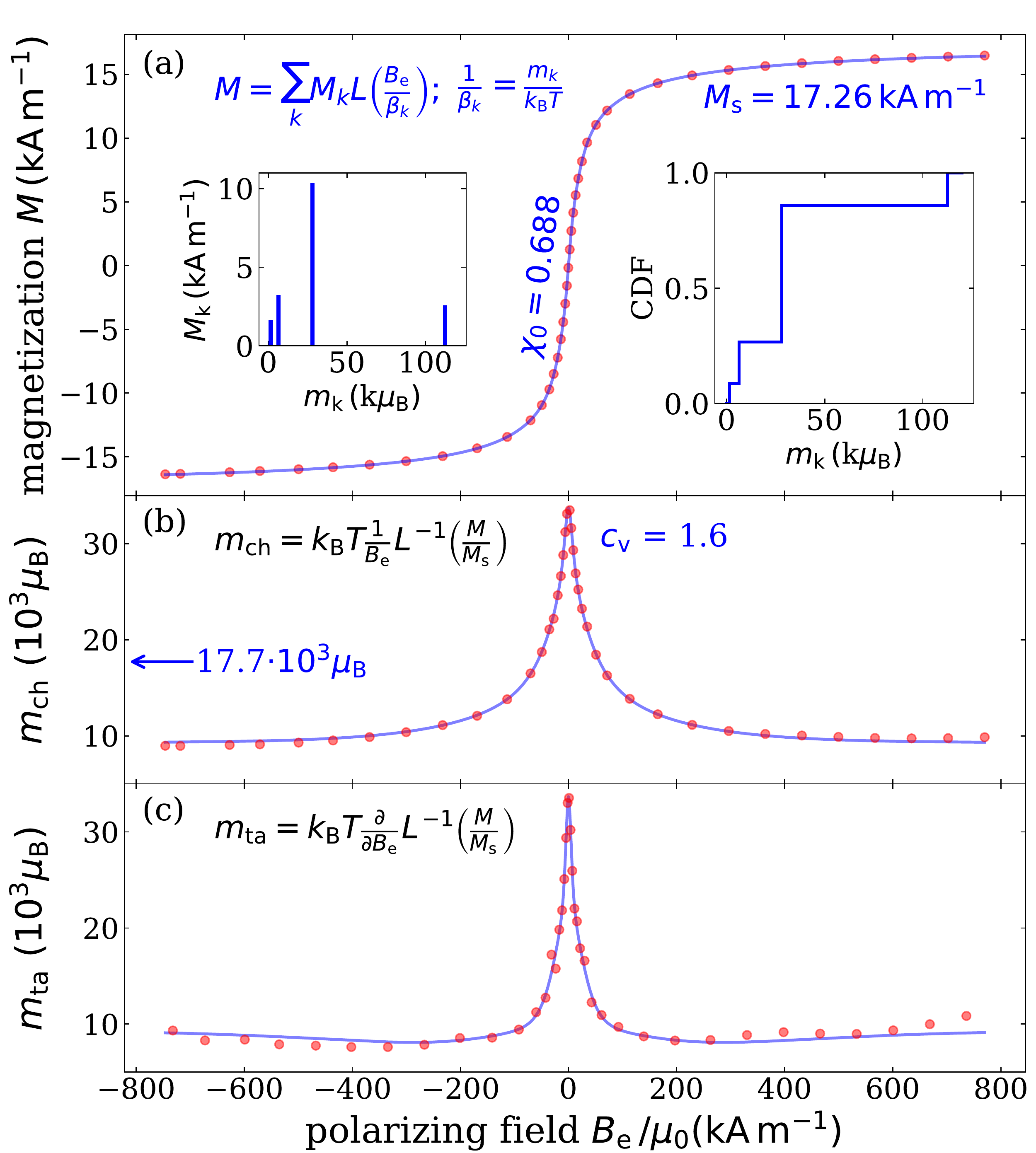}
\caption{
Examination of the ferrofluid EMG909. (a) The measured magnetization curve (red dots, only every 10th data point is shown) is fitted by the sum of four Langevin terms (solid blue line) indicated by the $M_k$. The corresponding $\beta_k$ yields the magnetic moment $m_k$. The respective $M_\mathrm{k}$ and $m_\mathrm{k}$ of the fitted function are shown in the left inset, the corresponding cumulative distribution function in the right inset. The resulting saturation magnetization $M_\mathrm{s}$ and initial susceptibility $\chi_\mathrm{0}$ are listed as well. (b) The effective magnetic moment $m_\mathrm{ch}$ obtained via the chord slope (eq.\,\ref{eq:m_ch}) from the data (red dots) and the fitting function (solid blue line). The corresponding $c_\mathrm{v}$ (eq.\,\ref{eqn:cv}) is listed. The blue arrow points to the value of the corresponding geometric mean (eq.\,\ref{eqn:gm}). (c) The effective magnetic moment $m_\mathrm{ta}$ obtained via the tangential slope (eq.\,\ref{eq:m_ta}) from the data (red dots) and the fitting function (solid blue line).}
\label{Figure1}
\end{figure}

We have measured the magnetization with a vibrating sample magnetometer utilizing a spherical sample holder described in Ref.\,\cite{Friedrich2012}. The result of this measurement is presented in Fig.\,\ref{Figure1}(a).
The "polarizing field" used for the horizontal axis is the field acting on a magnetic particle. It is determined by the external magnetic field $H_\mathrm{0}$ measured far from the spherical sample, and influenced by the homogeneous magnetization $M$ inside the sphere. We used the lowest order to determine the polarizing field, namely the Weiss correction $H_\mathrm{e}=H_\mathrm{i}+M/3$, where $H_\mathrm{i}$ is the magnetic field inside the spherical sample. A discussion of the Weiss correction and higher order corrections for the effective field $H_\mathrm{e}$ can be found in Ref.\,\cite{Ivanov2007}. For our magnetometer geometry, that correction term $M/3$ exactly cancels out the demagnetization factor provided by the spherical sample holder. This leads to $H_\mathrm{e}=H_\mathrm{0}$, and correspondingly $B_\mathrm{e}=B_\mathrm{0}$. Here $B_\mathrm{e}$ is the magnetic induction inside the virtual -- and hollow -- "Weiss sphere", which is responsible for the torque acting on the individual magnetic particle. In conclusion, in our case and in lowest order approximation the polarizing field $B_\mathrm{e}$ turns out to be the one measured far from our magnetized sphere $B_\mathrm{0}$, which is conveniently detected by a Hall probe. 

Note that this effective field correction is only a correction to lowest order for reasons of simplicity. According to eq.\ (20) of Ref.\,\cite{Ivanov2007}, the next correction term reads $\frac{1}{144} M_\mathrm{L} \frac{\mathrm{d}M_\mathrm{L}}{\mathrm{d}H_\mathrm{i}}$, which brings only a correction on a sub-percentage scale for the fluid presented here. For fluids with a higher magnetization, that term would become more important, i.e.\ the estimation of $B_\mathrm{e}$ presented here is only a lowest order approximation for sufficiently dilute solutions.

Neglecting this Weiss correction completely would have a measurable effect on the data examination: The initial slope of the magnetization curve would change by about 20\,\% (see eq.\,\ref{eq:ch0_chiL} below). This would lead to an error in the estimation of the arithmetic mean of the magnetic moments of the same order of magnitude \cite{Rehberg2019}. The harmonic mean, on the other hand, is determined from the slope of the magnetization at higher fields where the magnetization is almost saturated and the influence of the Weiss correction term decreases to zero. Thus, the harmonic mean would still be measured correctly, at least asymptotically for large fields.  A comparison of the two terms would then lead to a qualitatively wrong statement for a monodisperse fluid: It would be interpreted as a polydisperse one. In conclusion, neglecting the Weiss correction is clearly prohibited for the graphical granulometry \cite{Rehberg2019}.

Note that the resulting plot --- with the effective $B_\mathrm{e}$-field used for the x-axis --- is slightly different from the more common practice, where the inner magnetic field $H_\mathrm{i}$ is used for the horizontal axis of the magnetization curve. Our motivation to use $B_\mathrm{e}$ instead is the fact that magnetization curves in lowest order can be considered as a superposition of terms
\begin{equation*}\mathrm{L}\left(\frac{B_\mathrm{e}m}{k_\mathrm{B}T}\right), \mathrm{\ or\ equivalently \ } \mathrm{L}\left(\frac{\mu_\mathrm{0} H_\mathrm{e}m}{k_\mathrm{B}T}\right),  
\end{equation*} a fact which is used both for the best-fit curve in Fig.\,\ref{Figure1}(a), and the graphical granulometry provided in Fig.\,\ref{Figure1}(b) and Fig.\,\ref{Figure1}(c).  

The magnetization data shown in Fig.\,\ref{Figure1}(a) are obtained as the difference between the magnetization data from the filled and the empty sample holder, which is a lowest order correction for the magnetization of the sample holder. This is only a tiny correction of less than $0.1\%$ for the fairly concentrated magnetic fluid used here, it is nevertheless performed routinely.  

It turns out that the measured magnetization data can fairly accurately be represented by a superposition of four Langevin terms
\begin{equation}
\label{eqn:quad}
M(B_\mathrm{e})= \sum_{k=1}^{4} M_k \mathrm{L}\left(\frac{B_\mathrm{e}}{\beta_k}\right), \mathrm{\ with\ } \frac{1}{\beta_k}=\frac{m_k}{k_\mathrm{B}T}. 
\end {equation}
This $M(B_\mathrm{e})$ resulting from this "quad-disperse" distribution function provides a convenient fitting curve for the magnetization data, with the $M_k$ and $\beta_k$ as fit parameters.   The result is shown as a solid line in Fig.\,\ref{Figure1}(a). It serves primarily for giving a smooth and analytical representation of the data. 

The inset on the left hand side in Fig.\,\ref{Figure1}(a) is a graphical representation of the eight fit parameters. It might serve to give some feeling for the distribution function of the dipole strength in the polydisperse suspension. However, it should not be over-interpreted in this sense, other distribution functions would do the job almost as well, which is due to the ill-posed character of this inverse problem. More illustrations for this point are provided in Ref.\,\cite{Rehberg2019}. The corresponding cumulative distribution function (CDF) of the quad-disperse distribution is shown in the inset on the right hand side of Fig.\,\ref{Figure1}(a).

Some technical remarks about the fitting procedure have to be made. We found it useful to suppress negative values for the magnetic moments and the magnetization, which is conveniently done by squaring the corresponding terms in the fitting procedure, and finally taking the positive root of the resulting value. The method to obtain the eight fitting parameters of this particular representation contains four steps:
\begin{enumerate}[(i)]
\item Fit $M(B_\mathrm{e})$ with two parameters $M_1$ and $\beta_1$.
\item Keep $M_1$ and $\beta_1$, and allow for two additional fit parameters $M_2$ and $\beta_2$.
\item Keep the four parameters, and allow for two additional fit parameters $M_3$ and $\beta_3$.
\item Keep the six parameters, and allow for two additional fit parameters $M_4$ and $\beta_4$.
\end{enumerate}

It seems that in principle this list could be extended. In practice, we found for many of the fluids investigated so far that even six parameters are enough to describe the magnetization within the resolution of our data. On the other hand, we never needed more than eight parameters.

For the fitting procedure, we use a standard routine (named curve\_fit, from the package scipy.optimze \cite{scipy}) within the Python program minimizing the deviation between the data and the fitting function. Our program calls this function repeatedly, until a local minimum with respect to all the fit parameters is reached. It seems worth noticing that the ansatz given by eq.\,(\ref{eqn:quad}) is sufficiently simple and fast, so that the data processing can conveniently be done interactively. 
The fitting curve provides a noise-free representation of the data. It can be used to calculate the so-called Langevin susceptibility \begin{equation*}
\chi_\mathrm{L}=\frac{\mathrm{d}M}{\mathrm{d}H_\mathrm{e}}    
\end{equation*}
as the slope of the magnetization curve in its origin. From $\chi_\mathrm{L}$, the initial susceptibility is obtained as 
\begin{equation}
\chi_0=\frac{\mathrm{d}M}{\mathrm{d}H_\mathrm{i}}=\frac{\chi_\mathrm{L}}{1-\frac{\chi_\mathrm{L}}{3}}, 
\label{eq:ch0_chiL}
\end{equation}
which is provided in Fig.\,\ref{Figure1}(a). This number is an important macroscopic parameter for the hydrodynamic instability of this particular fluid investigated in Ref.\,\cite{Richter2020}. The other important characteristic number is the saturation magnetization of the fluid, which can be obtained from the fitting parameters as 
\begin{equation*}
M_\mathrm{s}=\sum_{k=1}^{4} M_k, 
\end{equation*}
a convenient way to extrapolate the data towards $B_\mathrm{e} \to \infty$. 

Fig.\,\ref{Figure1}(b) shows the effective magnetic moment $m_\mathrm{ch}$ obtained from the chord slope as 
\begin{equation}
m_\mathrm{ch}=k_\mathrm{B}T\frac{1}{B_\mathrm{e}}\mathrm{L}^{-1}\left( \ \frac{M}{M_\mathrm{s}}\right). 
\label{eq:m_ch}
\end{equation}
The red dots are obtained directly from the data, which does not cause any problem for large values of $B_\mathrm{e}$, but becomes difficult near the origin, where both $B_\mathrm{e}$ and $M$ are small. A careful calibration of the offset of the measured magnetization field is therefore crucial here. We do that by fitting a second order curve 
\begin{equation*}
    M=a_0 +a_1B +a_2B^2 \mathrm{\ ,with \ parameters \ } a_0,a_1,a_2
\end{equation*}
 to the data in a neighborhood of the polarizing field where the measured magnetization changes sign. 
The correction term for the magnetic field data is then obtained as
\begin{equation*}
    B_\mathrm{offset}=-\frac{a_1}{2a_2}\pm \sqrt{\frac{a_1^2}{4a_2^2}-\frac{a_0}{a_2}}.
\end{equation*}
The smaller one of these two solutions is subtracted from the measured values of the polarizing field to ensure that the average magnetization is very close to zero for $B_\mathrm{e}=0$. Note that this method does even work when the magnetization data contain hysteresis, provided that the magnetization curve is measured in both directions. This was indeed done for the measurement presented here.

The solid blue line in Fig.\,\ref{Figure1}(b), on the other hand, is obtained from the quad-disperse fitting function. Its analytic representation causes no problem with regard to taking the ratio of two small numbers. The solid line agrees fairly well with the discrete values obtained directly from the data. The comparison brings out a small asymmetry with respect to the sign of the polarizing field for the discrete data points, which the symmetric ansatz for the quad-disperse fitting function cannot produce.

The maximum value of the effective magnetic moment corresponds to the arithmetic mean of the magnetic moments of the particles $m_\mathrm{a}$, and the asymptotic value for large $B_\mathrm{e}$ to the harmonic mean $m_\mathrm{h}$, as explained in more detail in Ref.\,\cite{Rehberg2019}.

Their difference $m_\mathrm{a} - m_\mathrm{h}$ is a direct order parameter for the amount of polydispersity: It is zero for a monodisperse distribution and increases with the width of the distribution. In fact, this difference divided by the harmonic mean provides an estimator for the relative standard deviation (RSD, also called coefficient of variation $c_\mathrm{v}$). More precisely, we obtain the coefficient of variation as 
\begin{equation}
\label{eqn:cv}
c_\mathrm{v}=\sqrt{\frac{m_\mathrm{a}-m_\mathrm{h}}{m_\mathrm{h}}}.
\end{equation}
Its value is listed in the upper part of Fig.\,\ref{Figure1}(b).

Additionally, an estimator for the geometric mean is obtained by 
\begin{equation}
\label{eqn:gm}
m_\mathrm{g}=\sqrt{m_\mathrm{a}  m_\mathrm{h}}. 
\end{equation}
It is provided in the figure as well, and its value is indicated by the blue arrow pointing to the corresponding location on the vertical axis.

Note that the calculations leading to $c_\mathrm{v}$ and  $m_\mathrm{g}$ are only correct for certain distribution functions. Among those is the log-normal distribution, which seems to be the most prominent one assumed within the granulometric analysis of magnetization curves. It should be noted that the log-normal distribution tends to overestimate the fraction of large particles \cite{PSHENICHNIKOV1996}. An alternative comparison with a gamma distribution is presented in Ref.\,\cite{Rehberg2019}.  

The small differences between the data and the fitted curve in Fig.\,\ref{Figure1}(b) become more prominent in Fig.\,\ref{Figure1}(c), where the effective magnetic moment 
\begin{equation}
m_\mathrm{ta}=k_\mathrm{B}T\frac{\partial}{\partial B_\mathrm{e}}\mathrm{L}^{-1}\left( \ \frac{M}{M_\mathrm{s}}\right),
\label{eq:m_ta}
\end{equation}
obtained from the tangential slope,
is shown \cite{Rehberg2019}. Here the noise of the discrete data points is clearly larger. However, even here the signal/noise ratio seems good enough to extract the numbers for $m_\mathrm{a}$ and $m_\mathrm{h}$, and the corresponding guesses for the geometric mean $m_\mathrm{g}$ and the relative standard deviation $c_\mathrm{v}$.  
\section{Conclusion}
We propose the superposition of a few -- in our case four -- Langevin curves as an effective and precise model to describe the magnetization curve of real ferrofluids. The four effective magnetic moments in that fit should not be over-interpreted in the sense that they represent a distribution function of the magnetic moments. However, two characteristic values of that function, namely the arithmetic and the harmonic mean, can be safely read off from a plot of the slope of the inverse Langevin function of the magnetization data. This graphical magnetogranulometry is sufficient to get an estimate for the relative standard deviation of the distribution function of the magnetic moments.

\bibliographystyle{model1-num-names}


\bibliography{Rehberg_ICMF2019}

\begin{thebibliography}{6}
\expandafter\ifx\csname natexlab\endcsname\relax\def\natexlab#1{#1}\fi
\providecommand{\bibinfo}[2]{#2}
\ifx\xfnm\relax \def\xfnm[#1]{\unskip,\space#1}\fi
\bibitem[{Rehberg et~al.(2019)Rehberg, Richter, Hartung, Lucht, Hankiewicz, and
  Friedrich}]{Rehberg2019}
\bibinfo{author}{I.~Rehberg}, \bibinfo{author}{R.~Richter},
  \bibinfo{author}{S.~Hartung}, \bibinfo{author}{N.~Lucht},
  \bibinfo{author}{B.~Hankiewicz}, \bibinfo{author}{T.~Friedrich},
\newblock \bibinfo{title}{Measuring magnetic moments of polydisperse
  ferrofluids utilizing the inverse langevin function},
\newblock \bibinfo{journal}{Phys. Rev. B} \bibinfo{volume}{100}
  (\bibinfo{year}{2019}) \bibinfo{pages}{134425}.
\bibitem[{ICM(2019)}]{ICMF2019}
\bibinfo{title}{15th international conference on magnetic fluids},
  \bibinfo{howpublished}{\url{https://premc.org/conferences/icmf-magnetic-fluids/}},
  \bibinfo{year}{2019}.
\bibitem[{Völkel et~al.(2020)Völkel, Kögel, and Richter}]{Richter2020}
\bibinfo{author}{A.~Völkel}, \bibinfo{author}{A.~Kögel},
  \bibinfo{author}{R.~Richter},
\newblock \bibinfo{title}{Measuring the kelvin-helmholtz instability,
  stabilized by a tangential magnetic field},
\newblock \bibinfo{journal}{Journal of Magnetism and Magnetic Materials}
  \bibinfo{volume}{505} (\bibinfo{year}{2020}) \bibinfo{pages}{166693}.
\bibitem[{Friedrich et~al.(2012)Friedrich, Lang, Rehberg, and
  Richter}]{Friedrich2012}
\bibinfo{author}{T.~Friedrich}, \bibinfo{author}{T.~Lang},
  \bibinfo{author}{I.~Rehberg}, \bibinfo{author}{R.~Richter},
\newblock \bibinfo{title}{Spherical sample holders to improve the
  susceptibility measurement of superparamagnetic materials},
\newblock \bibinfo{journal}{Rev. Sci. Instr.} \bibinfo{volume}{83}
  (\bibinfo{year}{2012}) \bibinfo{pages}{045106--045106--7}.
\bibitem[{Ivanov et~al.(2007)Ivanov, Kantorovich, Reznikov, Holm,
  Pshenichnikov, Lebedev, Chremos, and Camp}]{Ivanov2007}
\bibinfo{author}{A.~O. Ivanov}, \bibinfo{author}{S.~S. Kantorovich},
  \bibinfo{author}{E.~N. Reznikov}, \bibinfo{author}{C.~Holm},
  \bibinfo{author}{A.~F. Pshenichnikov}, \bibinfo{author}{A.~V. Lebedev},
  \bibinfo{author}{A.~Chremos}, \bibinfo{author}{P.~J. Camp},
\newblock \bibinfo{title}{Magnetic properties of polydisperse ferrofluids: A
  critical comparison between experiment, theory, and computer simulation},
\newblock \bibinfo{journal}{Phys. Rev. E} \bibinfo{volume}{75}
  (\bibinfo{year}{2007}) \bibinfo{pages}{061405}.
\bibitem[{SciPy.org(2019)}]{scipy}
\bibinfo{author}{SciPy.org}, \bibinfo{title}{Scientific computing tools for
  python}, \bibinfo{howpublished}{\url{https://www.scipy.org}},
  \bibinfo{year}{2019}.

\end{thebibliography}
\end{document}